\documentclass[a4paper,11pt]{article}

\usepackage{lineno}

\usepackage[utf8]{inputenc}

\usepackage{hyperref} 
\usepackage{url}
\usepackage{graphicx}
\usepackage{amssymb}
\usepackage{amsmath}

\usepackage{multirow}

\usepackage{graphics}
\usepackage{natbib}
\graphicspath{{./FigureProc}}

\usepackage{xcolor}
\usepackage{textcomp}

\usepackage{orcidlink}

\usepackage{comment}
\usepackage[blocks]{authblk}

\title{Development of a novel compact and fast SiPM-based RICH detector for the future ALICE 3 PID system at LHC}

\author[a,1]{M.~N.~Mazziotta \orcidlink{0000-0001-9325-4672}}
\author[a]{A.~R.~Altamura \orcidlink{0000-0001-8048-5500}}
\author[a]{L.~Congedo \orcidlink{0000-0003-4536-4644}}
\author[a]{G.~De~Robertis \orcidlink{0000-0001-8261-6236}}
\author[b]{A.~Di~Mauro \orcidlink{0000-0003-0348-092X}}
\author[c]{J.~O. Guerra-Pulido}
\author[a]{F.~Licciulli \orcidlink{0000-0002-6955-0321}}
\author[a,d]{L.~Lorusso \orcidlink{0000-0002-2549-4401}}
\author[b]{P.~Martinengo \orcidlink{0000-0003-0288-202X}}
\author[a]{E.~Nappi \orcidlink{0000-0003-2080-9010}}
\author[a,d]{N.~Nicassio \orcidlink{0000-0002-7839-2951}}
\author[c]{G.~Pai\'c \orcidlink{0000-0003-2513-2459}}
\author[a,d]{G.~Panzarini \orcidlink{0000-0002-2586-1021}}
\author[a,d]{R.~Pillera \orcidlink{0000-0003-3808-963X}}
\author[a,d]{G.~Volpe \orcidlink{0000-0002-2921-2475}}

\affil[a]{Istituto Nazionale di Fisica Nucleare (INFN), Sezione di Bari, via Orabona 4, I-70126 Bari, Italy}
\affil[b]{CERN, the European Organization for Nuclear Research, Esplanade des Particules 1, 1211 Geneva, Switzerland}
\affil[c]{Instituto de Ciencias Nucleares, Universidad Nacional Autónoma de México, Mexico City (Mexico)}
\affil[d]{Dipartimento di Fisica dell'Universit\`a e del Politecnico di Bari, via Amendola 173, I-70126 Bari, Italy}
\affil[1]{Corresponding author: mazziotta@ba.infn.it}

\date{}
\begin{document}

\maketitle

\begin{abstract}
A dedicated R\&D is ongoing for the charged particle identification system of the \mbox{ALICE 3} experiment proposed for the LHC Run 5 and beyond.
One of the subsystems for the high-energy charged particle identification will be a Ring-Imaging Cherenkov (RICH) detector. 
The possibility of integrating Cherenkov-based charged particle timing measurements is currently under study. 
The proposed system is based on a proximity-focusing RICH configuration including an aerogel radiator separated from a SiPM array layer by an expansion gap. A thin high-refractive index window of transparent material, acting as a second Cherenkov radiator, is glued on the SiPM array to enable time-of-flight measurements of charged particles by exploiting the yield of Cherenkov photons in the thin window. 
We assembled a small-scale prototype instrumented with different Hamamatsu SiPM array sensors with pitches ranging from 1 to 3 mm, readout by custom boards equipped with the front-end Petiroc 2A ASICs to measure charges and times. 
The primary Cherenkov radiator consisted of a 2 cm thick aerogel tile, while various window materials, including SiO$_2$ and MgF$_2$, were used as secondary Cherenkov radiators.
The prototype was successfully tested in a campaign at the CERN PS T10 beam line with pions and protons. This paper summarizes the results achieved in the 2023 test beam campaign.
\end{abstract}

\section{Introduction}
\label{sec:intro}

The excellent properties of silicon photomultipliers (SiPMs) in terms of photon detection efficiency (PDE) and single photoelectron time resolution suggest the possibility of designing a  compact and fast Ring-Imaging Cherenkov (RICH) system integrating Cherenkov-based timing measurements.
A dedicated R\&D is in progress for the optimization of the future charged particle identification system for the ALICE 3 experiment, planned as a follow-up to the current ALICE apparatus for LHC Runs 5 and 6~\cite{ALICE3Loi}.

The proposed system is based on a proximity-focusing RICH configuration using an aerogel tile as Cherenkov radiator separated from a SiPM array layer by an expansion gap. A thin high-refractive index slab of transparent material, acting as a second Cherenkov radiator for the impinging charged particles, is directly coupled to the SiPMs to achieve an efficient and precise charged particle timing.

A prototype has been successfully assembled and tested at the CERN-PS T10 in October 2022~\cite{Nicassio_proceeding_IWASI} and 2023. In this work the results achieved in the 2023 test beam campaign are presented.

\section{Beam test set-up}
\label{sec:setup}

\begin{figure}
\centering
\includegraphics[height=0.43\textheight,width=0.95\linewidth]{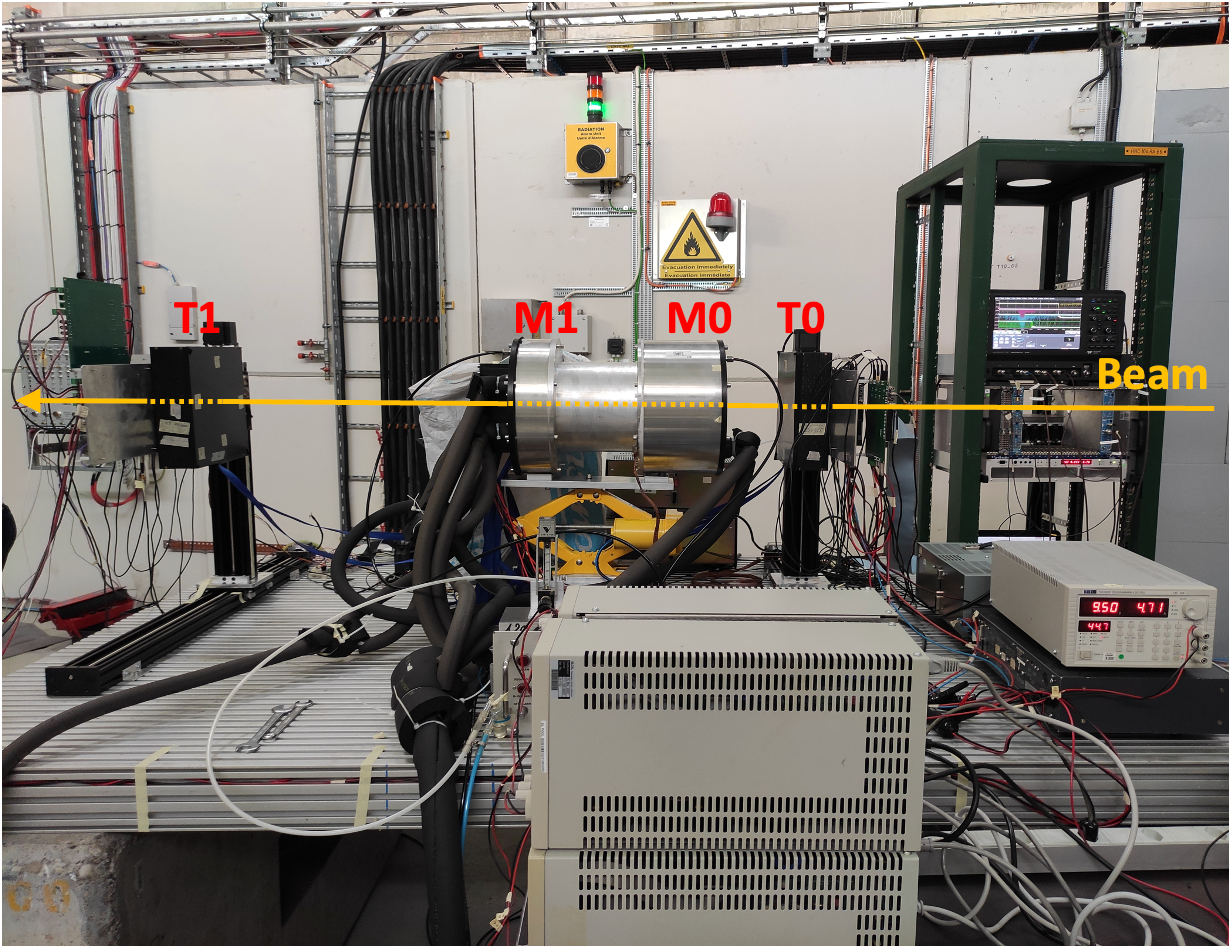}
\caption{Beam test set-up at CERN PS T10 line on Oct, 2023. The beam enters from the right side. The black boxes upstream and downstream the set-up include thin plastic scintillator tiles and two X-Y fiber tracker modules (T0 and T1)~\cite{MAZZIOTTA2022167040}.}
\label{fig:setup}
\end{figure}

\begin{figure}
\centering
\includegraphics[height=0.4\textheight,width=0.9\linewidth]{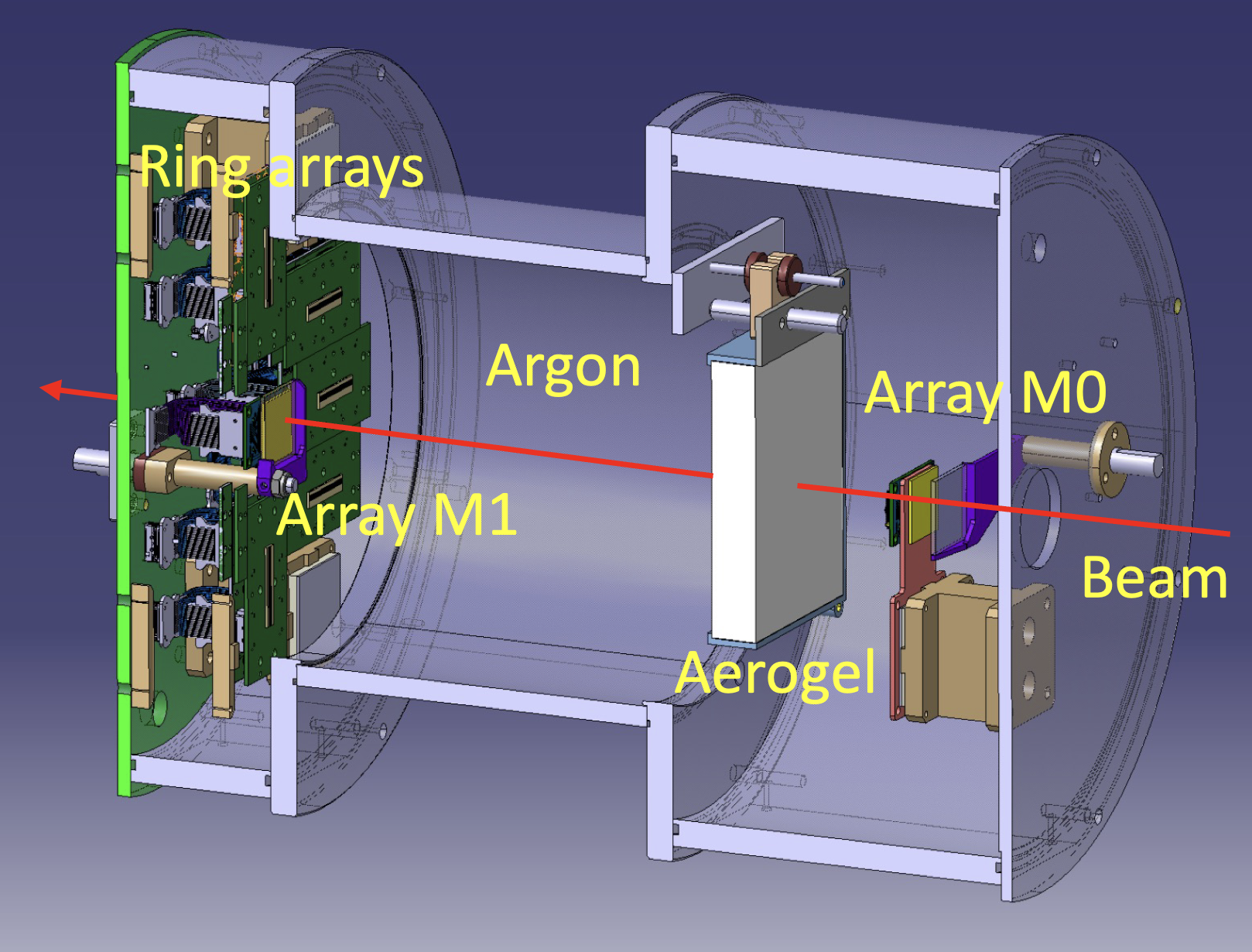}
\caption{CAD view of the RICH cylinder with the central upstream SiPM array M0, the aerogel tile, the SiPM linear ring arrays and the central downstream SiPM array M1. }
\label{fig:proto}
\end{figure}

The overall set-up at the CERN-PS T10 beam line is shown in Fig.~\ref{fig:setup}.
The tested prototype shown in Fig.~\ref{fig:proto} consists of a cylinder housing a 2 cm thick aerogel from Aerogel Factory \& Co., Ltd. with a refractive index of $n$ = 1.03 at 400~nm wavelength~\cite{AnnaRita_proceeding_IWASI}. The Cherenkov radiator is 23~cm far away from a photon detector layer consisting of eight HPK S13552 SiPM arrays~\cite{s13552}, the centers of which are assembled along a circumference of about 6.5~cm of radius.

Two additional 8$\times$8 HPK S13361-3075AE-08 SiPM arrays (hereafter M0 and M1) are placed along the beam line to perform charged particle timing measurements. A 1~mm thick window, acting as a second Cherenkov radiator, is directly coupled to each of those arrays to study the charged particle timing thanks to the prompt production of Cherenkov photons in the window. 
A 1 mm thick MgF$_{2}$ ($n$ = 1.38 at 400~nm) window is glued on M0, while a 1 mm thick SiO$_{2}$ ($n$ = 1.47 at 400~nm) radiator is glued on M1. These radiators ensure a Cherenkov photon emission starting from momenta lower than 1 GeV/c for all the particle species of interest.

\begin{figure}
\centering
\includegraphics[height=0.27\textheight]{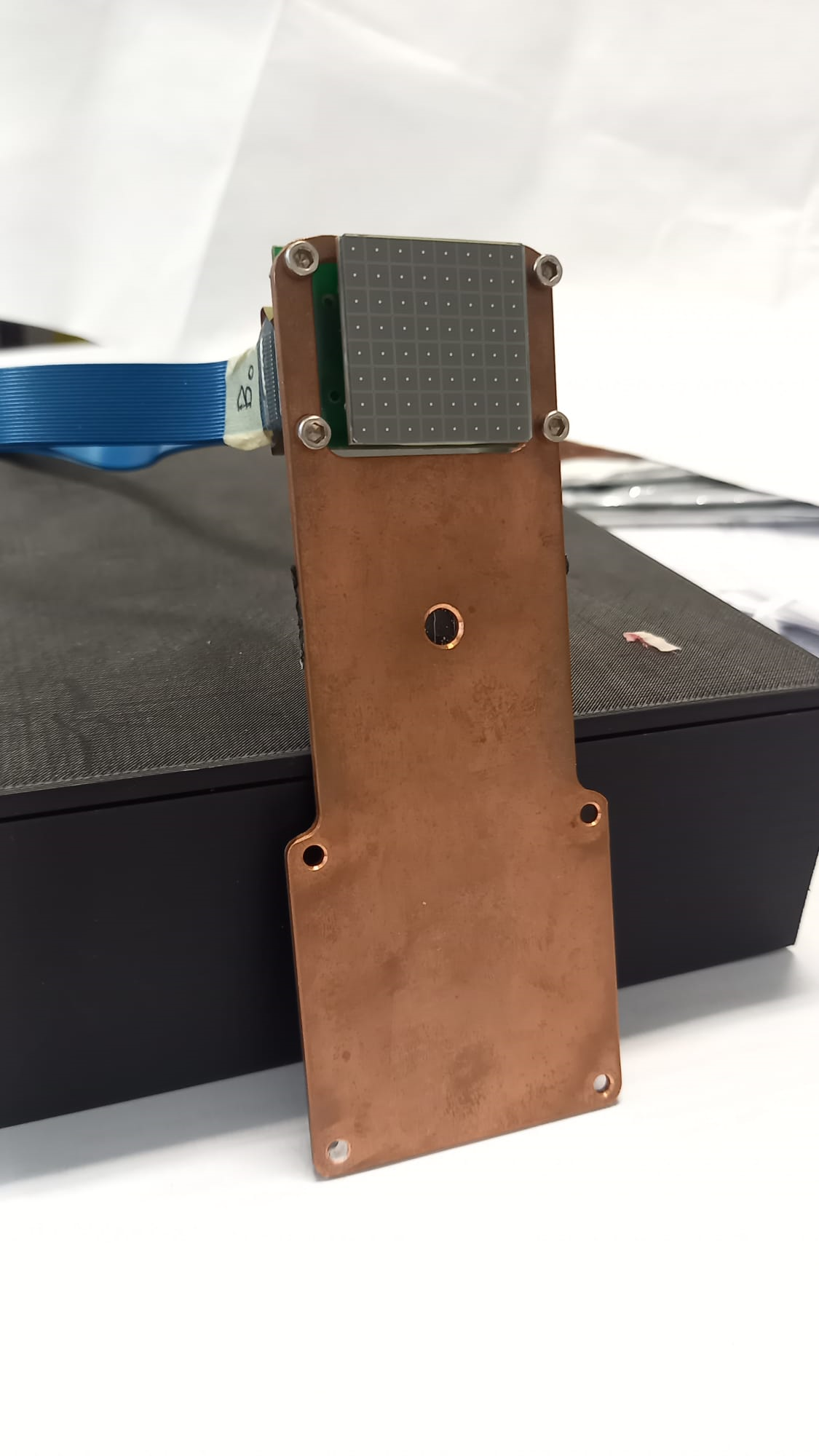}
\hspace{0.5cm}
\includegraphics[height=0.27\textheight,width=0.49\linewidth]{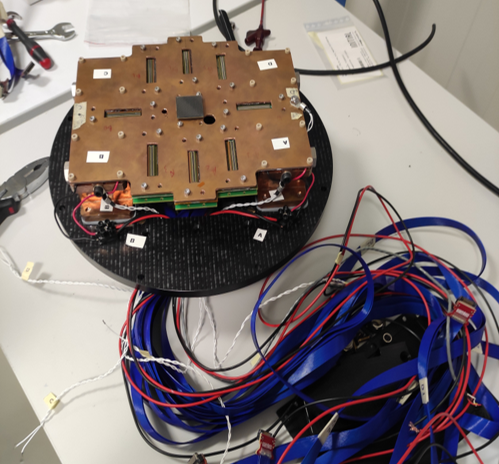}
\caption{Left panel: upstream S13361-3075AE-08 SiPM array coupled with a 1~mm thick MgF$_2$ window. Right panel: downstream photon detector plate with eight S13552 SiPM arrays and a central S13361-3075AE-08 SiPM array coupled with a 1~mm thick SiO$_2$ window.}
\label{fig:array}
\end{figure}

The arrays in the cylinder are mounted on copper plates, as shown in Fig.~\ref{fig:array}, and cooled down to -5$^\circ$C using water chillers and Peltier cells. The cylinder is flushed with Ar to keep the relative humidity $<$ 2\%. 

Ancillary systems include scintillator tiles for triggering and two X-Y fiber tracker modules, T0 upstream and T1 downstream, still equipped with S13552 arrays~\cite{MAZZIOTTA2022167040} operated at room temperature~(see Fig. \ref{fig:setup}). Ring arrays, central arrays and tracker arrays are operated at an over-voltage of 4.6, 6.3 and 3.0 V, respectively.

Each S13552 128-channels SiPM array is bonded on one side of a printed circuit board (PCB) and all channels are routed to four Samtec LSHM-120 multi-channel connectors~\cite{lshm} bonded on the other side of the PCB. Then, each S13552 SiPM array is plugged to a mezzanine PCB with four LSHM-120 connectors on one side to match the SiPM carrier board, and one LSHM-120  connector on the other side. In this way,
each S13552 is readout with a single 32-channel ASIC grouping four adjacent channels, resulting in a readout pitch of 1 mm.

Each S13361-3075AE-08 64-channels SiPM array is plugged with its two Samtec ST4-40-1.00-L-D-P-TR connectors~\cite{s13661,st4} to a custom mezzanine PCB that hosts two SS4-40-3-00-L-D-K connectors~\cite{ss4} on one side and two LSHM-120 connectors on the other side.

The analog SiPM signals are read-out by custom front-end boards (FEB) in master-slave mode~ \cite{MAZZIOTTA2022167040,Roberta_proceeding_IWASI} by means of 1 meter long high speed 50 Ohm multi-channels HLCD SAMTEC cables~\cite{hlcd}. Then each HLCD with 32 anlog signals is read-out with a 32-channel Petiroc 2A ASIC (37 ps LSB TDC, 10 bit ADC)~\cite{petiroc2a}.
Each FEB hosts four Petiroc 2A ASICs, a CAEN A7585D SiPM voltage module~\cite{a7585} and a Kintex-7 FPGA mounted on a Mercury+ KX2 module~\cite{fpgakx2} to configure both the ASICs and the trigger and to manage the data acquisition (DAQ).

The SiPM carrier boards are equipped with 1-wire digital temperature sensors DS18B20~\cite{ds18b20} to monitor their temperatures, which are read-out by the FPGA. Further analog temperature sensors Thermistor NTC 10k~\cite{ntc10k} are assembled on the copper plates to monitor their temperatures by using Raspberry Pi 3 with ADS1115 16-Bit ADCs \cite{ads1115}. In addition, humidity sensors SHT31 \cite{sht31} have been used to monitor the room and cylinder humidity.

\section{Data analysis and results}
\label{sec:data}

For each dataset we evaluated the net ADC counts of individual channels by subtracting the corresponding pedestal values. Then, we identified the peaks in the net ADC charge distribution corresponding to the number of detected photoelectrons (PEs) and finally we calculated the calibration ADC-to-PE constants. 
We also calculated the number of charged particle cluster PEs in the two fiber tracker planes T0 and T1 and the two central SiPM arrays M0 and M1 by merging the measured charge of geometrically adjacent channels.
A time calibration was performed at channel-by-channel level by using the TDC values. We first removed  any time offset evaluating the time delay for every SiPM channel (time-of-flight, routing, cabling, etc.). Then, a time walk correction was performed according to the observed number of PEs.

The event selection was performed requiring hits in an about 8$\times$9 mm$^2$ fiducial area of both the two fiber tracker planes and the two central SiPM arrays.
The beam direction was assumed along the Z-axis. The incoming charged particle directions in the X-Z and Y-Z views, respectively, 
were evaluated with a straight line fit by means of the hit positions in the central and the tracker SiPM arrays in each of the two views.

We assumed that all the hits in the ring arrays were candidate Cherenkov photons emitted in aerogel. For each candidate photon, the hypothetical emission direction was calculated by geometric backpropagation from the SiPM ring arrays to the extrapolated charged particle X-Y coordinates in the median plane of the aerogel tile.
In the backpropagation procedure, corrections by Snell’s law were applied to account for refraction at the interfaces between different media (the refractive indices of the different materials are assumed at 400 nm). 
The Cherenkov angle was evaluated as the angle between the track and the candidate Cherenkov photon directions.

\begin{figure}
\centering
\includegraphics[height=0.2\textheight, width=0.49\linewidth]{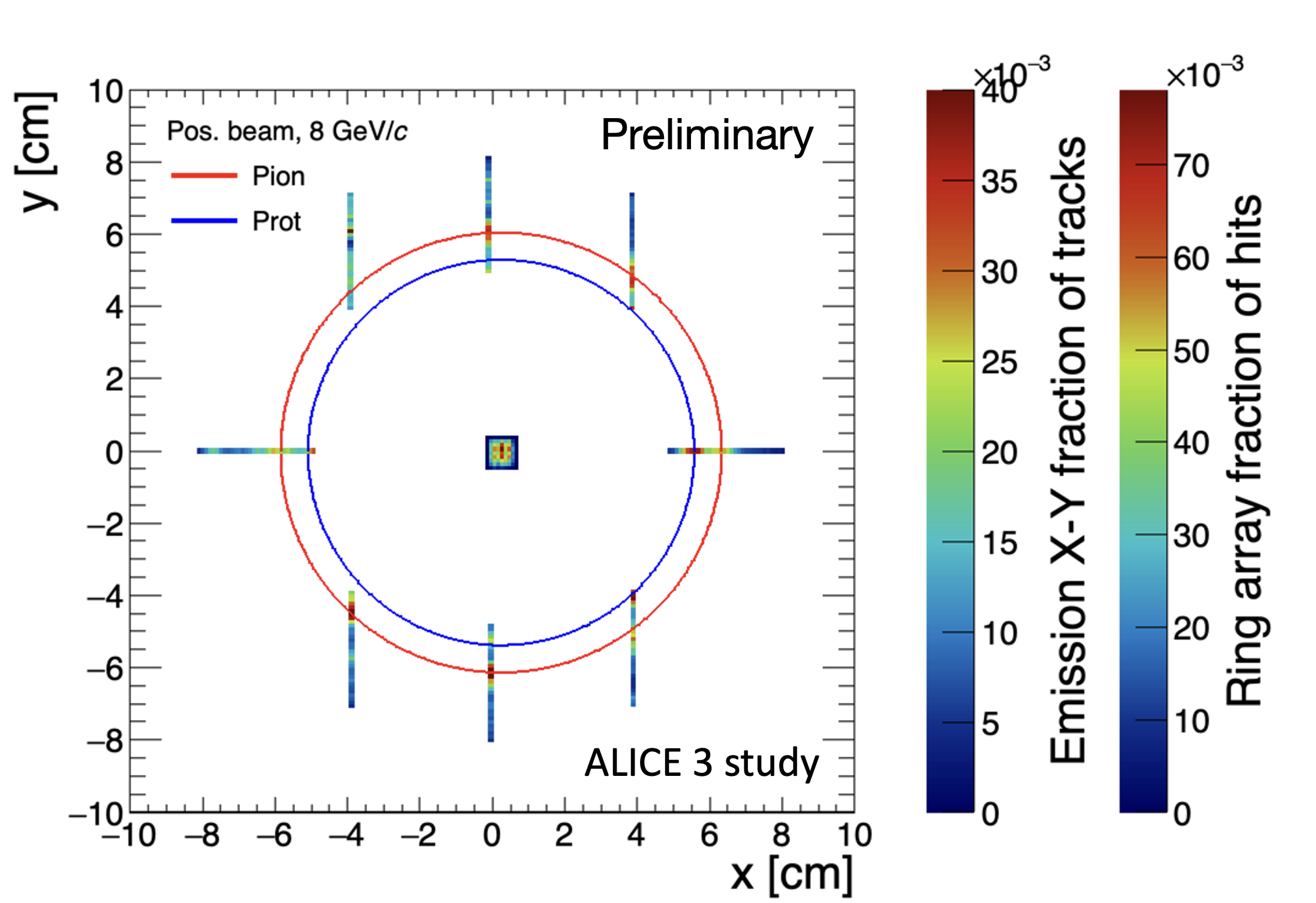}
\includegraphics[height=0.2\textheight, width=0.49\linewidth]{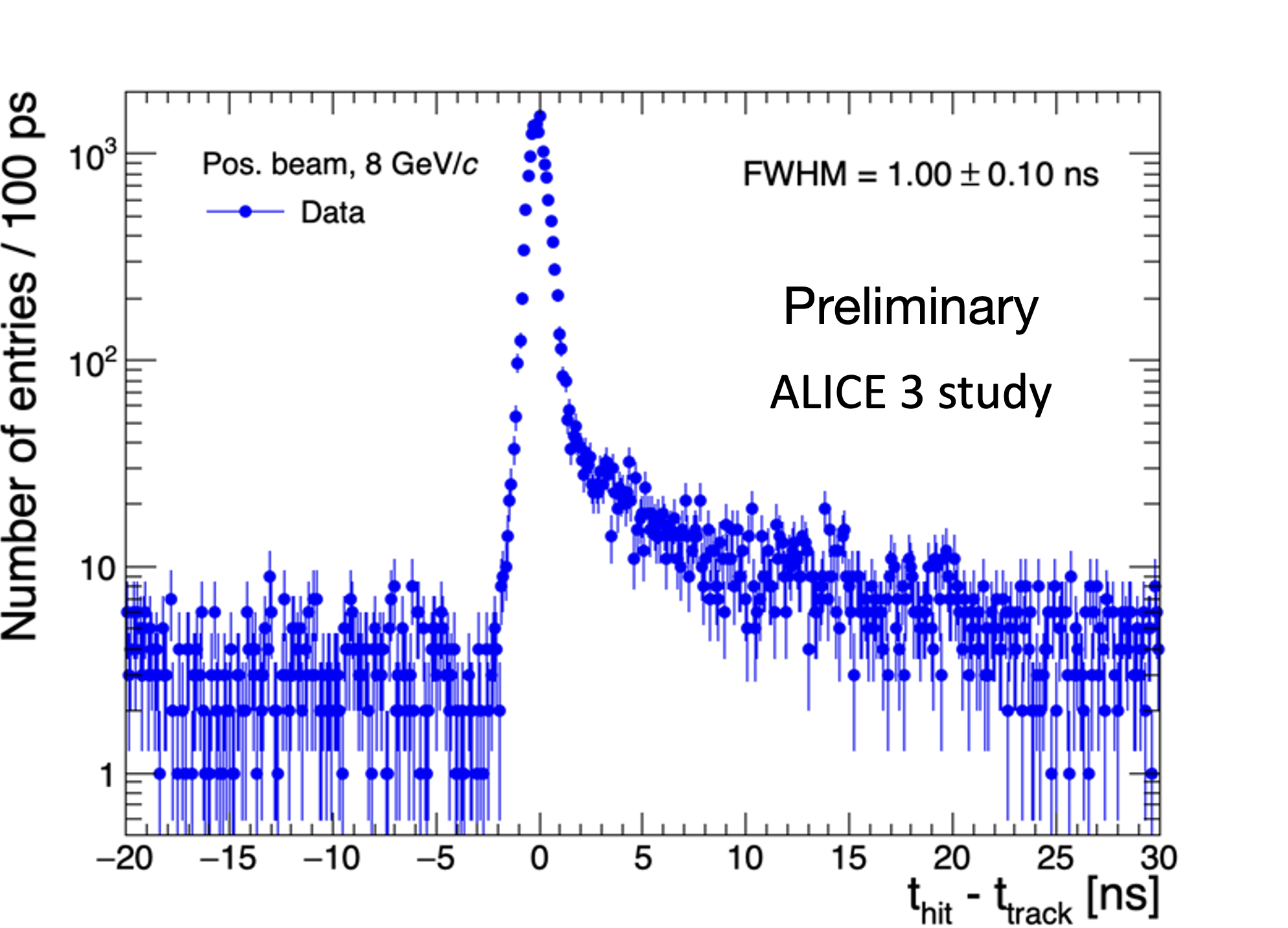}
\includegraphics[height=0.2\textheight,width=0.49\linewidth]{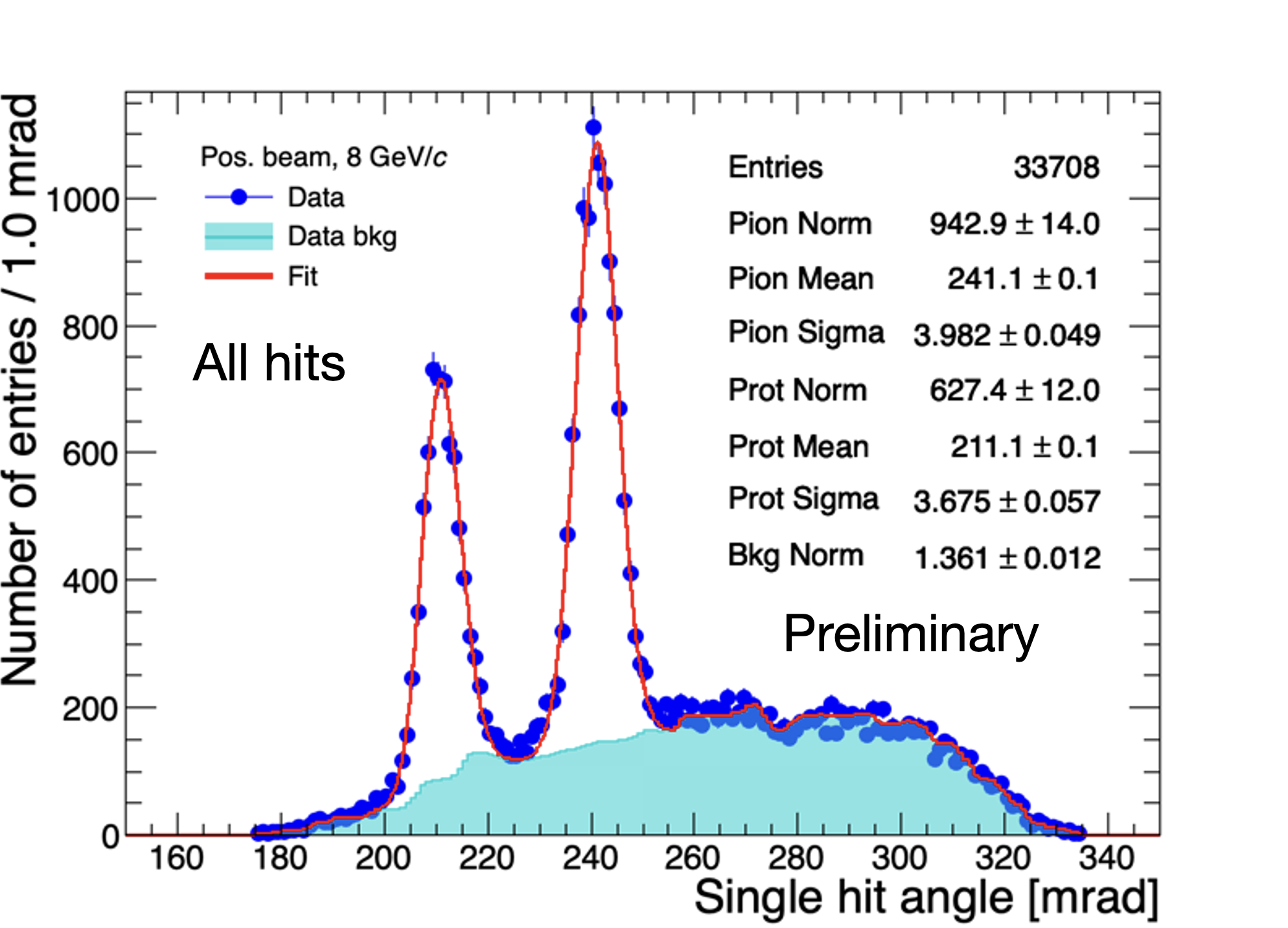}
\includegraphics[height=0.2\textheight, width=0.49\linewidth]{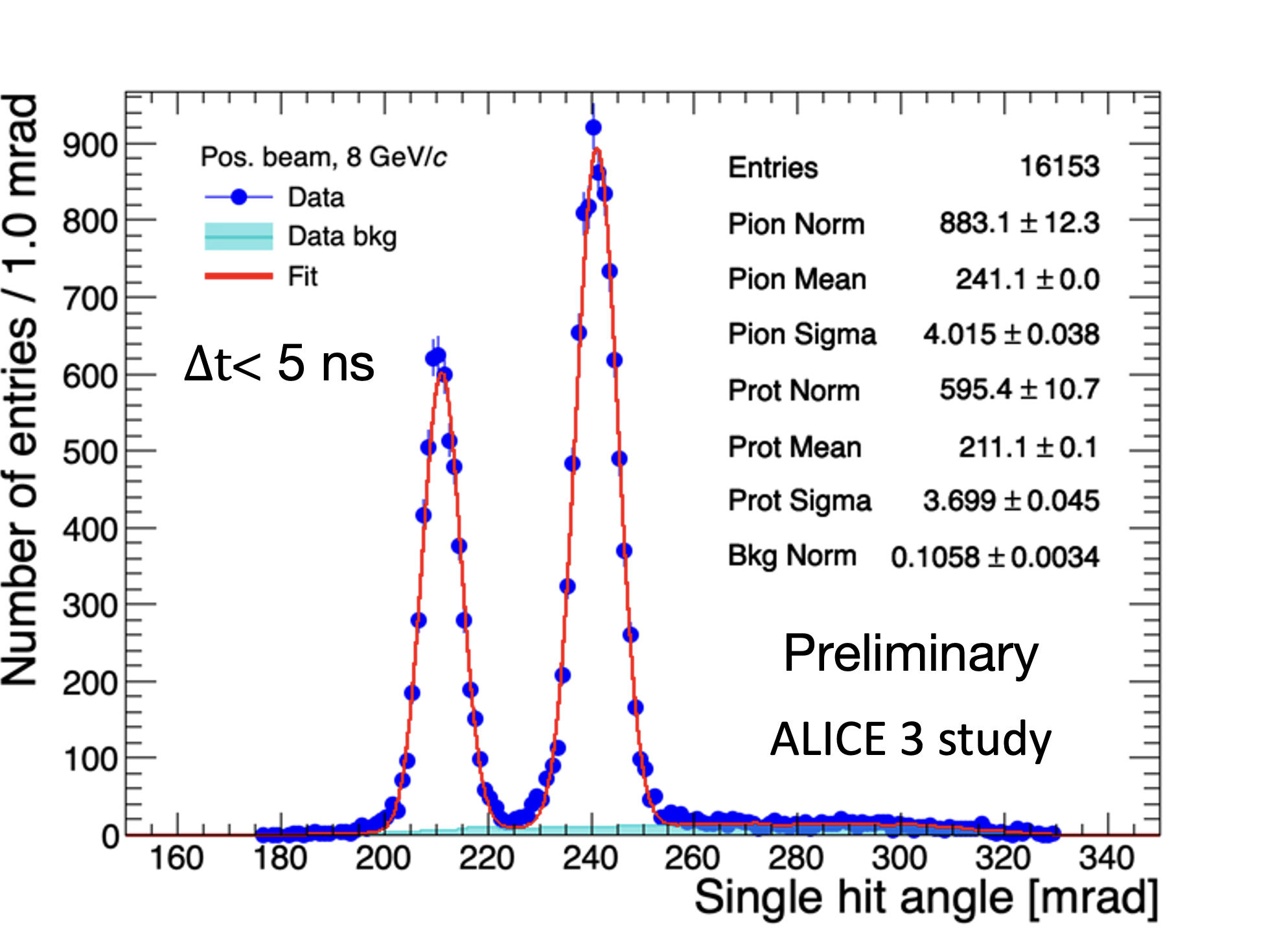}
\caption{Top left: Spatial distribution of the extrapolated X-Y emission coordinates in aerogel and of the hits in the ring arrays. Top right: Distribution of the relative timing between the array cell firing time $\text{t}_{\text{hit}}$ and the firing time $\text{t}_{\text{track}}$ of the M1 cell with maximum number of PEs. Bottom left: Raw single photon Cherenkov angle distribution. Bottom right: Corresponding angular distribution requiring $|\text{t}_{\text{hit}} - \text{t}_{\text{track}}|<$ 5~ns. The results refer to measurements with the positive beam at 8 GeV/$c$ momentum.}
\label{fig:st2023_rich}
\end{figure}

\begin{figure}
\centering
\includegraphics[height=0.2\textheight, width=0.49\linewidth]{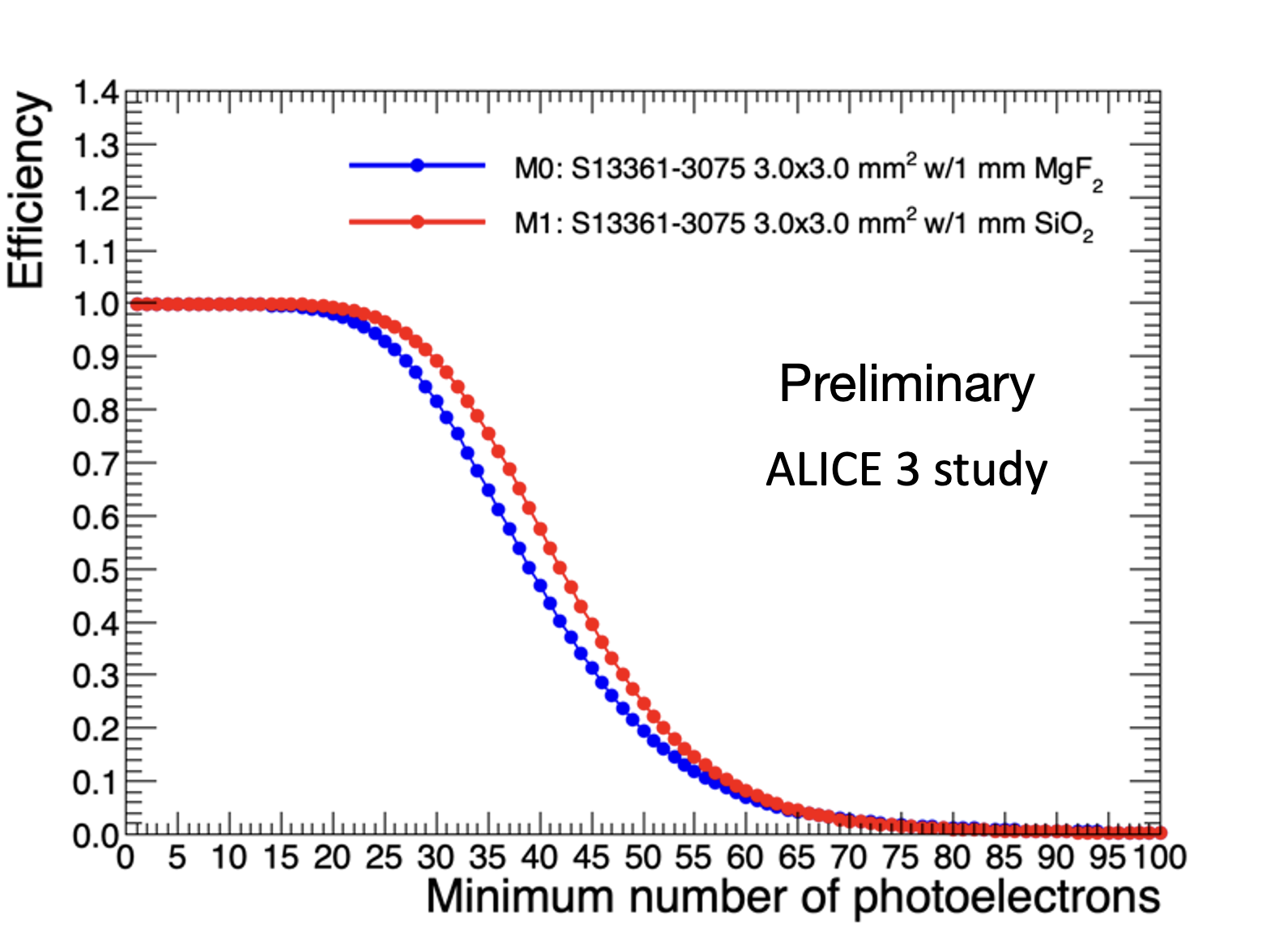}
\includegraphics[height=0.2\textheight, width=0.49\linewidth]{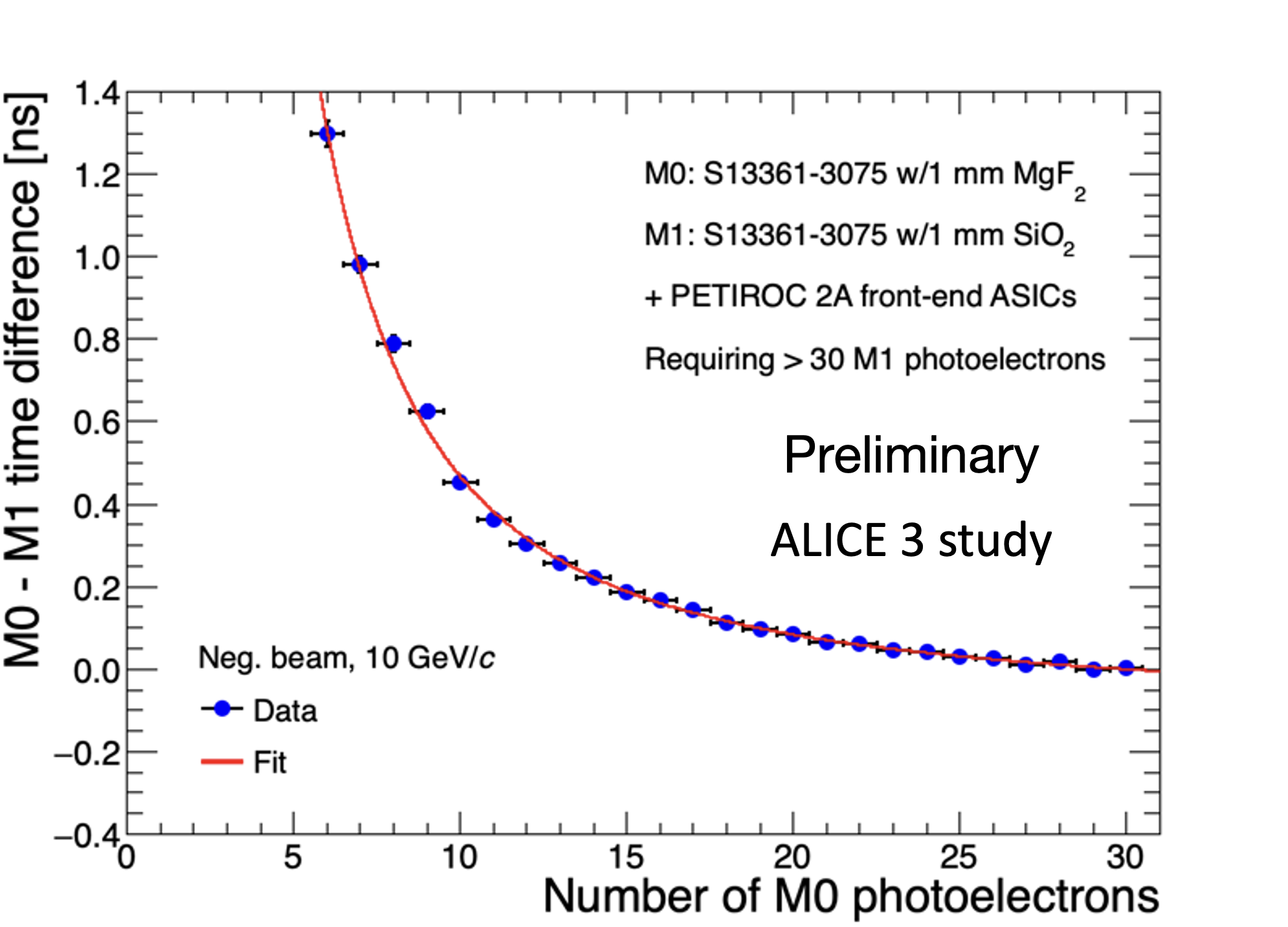}
\includegraphics[height=0.2\textheight,width=0.49\linewidth]{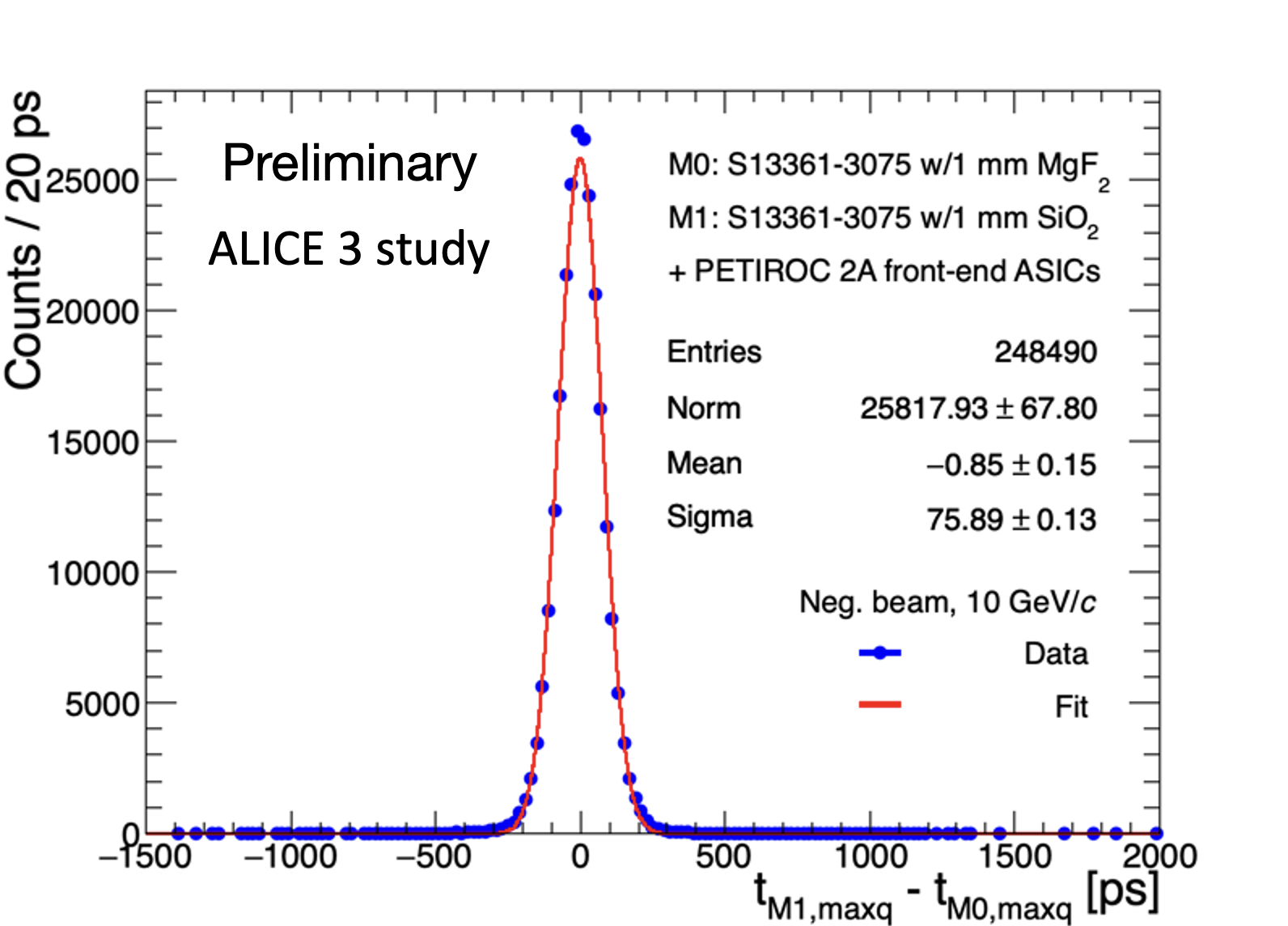}
\includegraphics[height=0.2\textheight, width=0.49\linewidth]{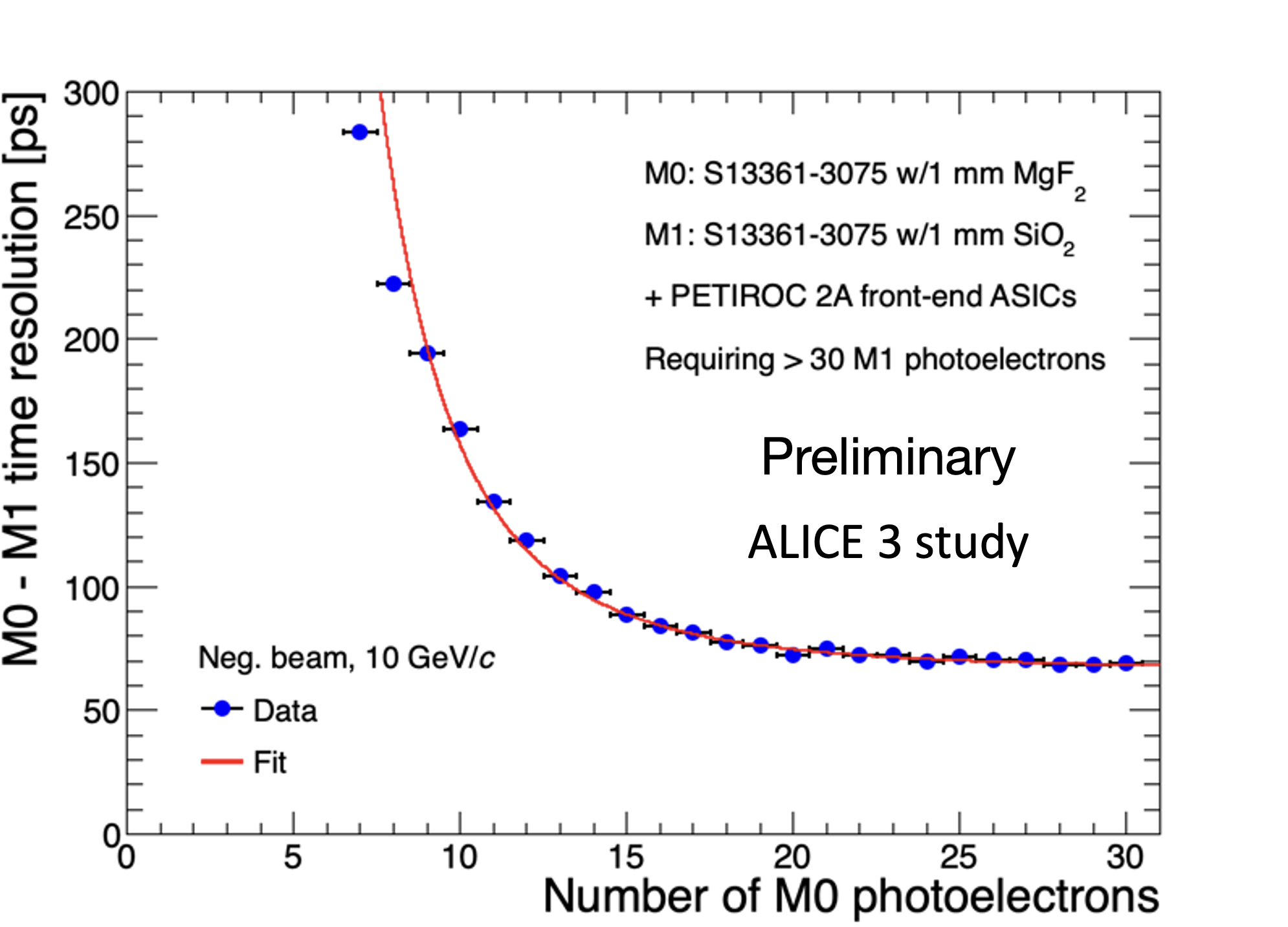}
\caption{Top left: M0 (blue points) and M1 (red points) charged track detection efficiency vs minimum number of required cluster PEs. Top right: Time walk correction for M0 channels. Bottom left: Distribution of the relative time of the M0 and M1 channels with maximum number of PEs. Bottom right: Time resolution as a function of M0 number of PEs requiring $>$ 30 PEs in M1. The results refer to measurements with the negative beam at 10 GeV/$c$ momentum.}
\label{fig:st2023_tof}
\end{figure}

Uncorrelated background counts in the Cherenkov angle distribution were mainly due to the random firing of ring array channels due to the SiPM dark counts rate (DCR) and electronics noise. 
A template of uncorrelated background hits was calculated with dedicated runs adding a 10 $\mu$s delay in the trigger system and by running the same data analysis chain discussed above. Then, a DCR hit suppression was performed requiring relative timing $|\text{t}_{\text{hit}} - \text{t}_{\text{track}}| < $ 5 ns, 
where $\text{t}_{\text{hit}}$ is the measured time in the ring arrays and $\text{t}_{\text{track}}$ is the measured time for the channel with the maximum number of PEs in M1.

The top left panel of Fig.~\ref{fig:st2023_rich} shows the measured spatial distributions of the extrapolated X-Y emission points in aerogel and of the hits in the ring arrays for the positive charged beam at 8 GeV/$\it{c}$ mainly made of pions and protons. The expected clusters due to Cherenkov photons from pions and protons are clearly visible. The corresponding Cherenkov rings centered in the most probable X-Y emission point are also shown.
The top right panel of Fig.~\ref{fig:st2023_rich} shows the distribution of the relative time between $\text{t}_{\text{hit}}$  and $\text{t}_{\text{track}}$.
The core of that distribution has a Gaussian shape with a sigma of approximately 400 ps.

The distribution of the reconstructed Cherenkov emission angle at single hit level (not corrected for the acceptance) for the positive beam at 8 GeV/$\it{c}$ is shown in the bottom left panel of Fig.~\ref{fig:st2023_rich}. The corresponding distribution filtered by selecting only hits such that $|\text{t}_{\text{hit}} - \text{t}_{\text{track}}| < 5$ ns is shown in the bottom right panel of the same figure.
The expected peaks due to Cherenkov photons from pions and protons are clearly visible. The distributions are fitted with the sum of Gaussians corresponding to the signal peaks and a template background distribution, as discussed above. We measured single hit Chrenkov angle resolutions better than 4 mrad for both pions and protons.

Finally, charged particle timing measurements were performed by evaluating the relative time of track clusters in M0 and M1. The timing results reported here were achieved with negative beam at 10 GeV/c momentum.
The top left panel of Fig.~\ref{fig:st2023_tof} shows the M0 and M1 efficiencies as function of the minimum number of required PEs in the charged particle clusters. Charged particle detection efficiencies close to $\approx100$\% were achieved requiring up to above 15 PEs in both M0 and M1.
The top right panel shows the relative time of M0 with respect to M1 as a function of the observed number of M0 PEs requiring at least 30 PEs in M1. The resulting curve represents the time function used to correct the time walk effect of M0. Similarly, we evaluated the time function to correct the time walk of M1~(not shown). Then we evaluated the time resolution between M0 and M1. The bottom left panel of Fig.~\ref{fig:st2023_tof} shows the time difference between M1 and M0, while the bottom right panel shows the time resolution of M0 as a function of PEs by requiring at least 30 PEs in M1. We measured a time resolution down to 75 ps on the relative time between M1 and M0, corresponding to about 50 ps or better at single SiPM channel accounting both the electronics and the intrinsic SiPM contributions.

\section{Conclusions}
\label{sec:conc}

We measured a single photon angular resolution close to Cherenkov angle saturation in aerogel better than 4 mrad. Excellent background suppression was achieved using a good hit-track time matching.
A charged particle detection efficiency close to $\approx100\%$ was achieved with a 1 mm thick window of MgF$_{2}$ and SiO$_{2}$ glued onto the SiPM arrays. 
The measured time resolution on the relative time between M1 and M0 was $\approx$ 75 ps, corresponding to about 50 ps or better at single SiPM channel accounting both the electronics and the intrinsic SiPM contributions.
The results are consistent with the state-of-the-art ALICE 3 RICH requirements, making the proposed SiPM-based PID system also attractive for future high-energy physics experiments and for space applications. 

\section*{Acknowledgments}
The authors would like to thank the INFN Bari staff for its contribution to the procurement and to the construction of the prototype. In particular, we thank D. Dell'Olio, M. Franco, N. Lacalamita, F. Maiorano, M. Mongelli, M. Papagni, C. Pastore and R. Triggiani for their technical support.

The authors also acknowledge the CERN team for facilities, organization of the beam test and support throughout all the beam test duration.

The research leading to these results received  funding from the European Union's Horizon Europe research and Innovation programme under grant agreement No. 101057511 (EURO-LABS).


\bibliographystyle{unsrt}
\bibliography{Proceeding.bib}

\end{document}